%% file: iclr2024_conference.tex
\documentclass{article} 
\usepackage{iclr2024_conference,times}

\input{math_commands.tex}

\usepackage{hyperref} 
\usepackage{url}
\usepackage{amssymb}
\usepackage{graphicx}
\usepackage{caption} 
\usepackage{float} 
\usepackage{wrapfig}
\usepackage{cleveref}
\usepackage{amsbsy}

\title{A Denoising VAE for Intracardiac Time Series in Ischemic Cardiomyopathy}

\author{Samuel Ruipérez-Campillo
\thanks{ \hspace{0.1em}First Author's correspondence address: ETH Zürich, CAB G15.2, Universitätstrasse 6, 8092, Switzerland. \\ \phantom{aa}\hspace{0.2em} \textsuperscript{\dag} Co-senior authors.}\\
Dept. of Computer Science\\
ETH Zurich \\
{\scriptsize \texttt{samuel.ruiperezcampillo@inf.ethz.ch}} \\
\And
Alain Ryser, Thomas M. Sutter, \\
Dept. of Computer Science\\
ETH Zurich \\
{\scriptsize \texttt{\{alain.ryser,thomas.sutter\}@inf.ethz.ch}}
\And
{\small Ruibin Feng, Prasanth Ganesan, Brototo Deb, Kelly A. Brennan, Maxime Pedron, Albert J. Rogers}\\
Dept. of Medicine\\
Stanford University\\
{\scriptsize \texttt{\{ruibin, prash030, bdeb7, kbrenn, mpedron, rogersaj\}@stanford.edu}}
\And
Maarten Z.H. Kolk, Fleur V.Y. Tjong, \\
Dept. of Cardiology\\
Amsterdam University\\
{\scriptsize \texttt{\{m.kolk,f.v.tjong\}@amsterdamumc.nl}}
\And
Sanjiv M. Narayan\textsuperscript{\dag}, \\
Dept. of Medicine\\
Stanford University\\
{\scriptsize \texttt{sanjiv1@stanford.edu}}
\And
Julia E. Vogt\textsuperscript{\dag} \\
Dept. of Computer Science\\
ETH Zurich\\
{\scriptsize \texttt{julia.vogt@inf.ethz.ch}}
}

\iclrfinalcopy 
\begin{document}

\maketitle

\begin{abstract}
In the field of cardiac electrophysiology (EP), effectively reducing noise in intra-cardiac signals is crucial for the accurate diagnosis and treatment of arrhythmias and cardiomyopathies. However, traditional noise reduction techniques fall short in addressing the diverse noise patterns from various sources, often non-linear and non-stationary, present in these signals. This work introduces a Variational Autoencoder (VAE) model, aimed at improving the quality of intra-ventricular monophasic action potential (MAP) signal recordings. By constructing representations of \textit{clean} signals from a dataset of 5706 time series from 42 patients diagnosed with ischemic cardiomyopathy, our approach demonstrates superior denoising performance when compared to conventional filtering methods commonly employed in clinical settings. We assess the effectiveness of our VAE model using various metrics, indicating its superior capability to denoise signals across different noise types, including time-varying non-linear noise frequently found in clinical settings. These results reveal that VAEs can eliminate diverse sources of noise in single beats, outperforming state-of-the-art denoising techniques and potentially improving treatment efficacy in cardiac EP.

\end{abstract}

\section{Introduction}
Arrhythmias and cardiomyopathies  pose significant challenges in cardiac healthcare. These conditions not only impact patient well-being but also complicate the clinical approach to treatment \citep{benjamin2018heart, gopinathannair2015arrhythmia}. Ablation therapy, a key intervention for many arrhythmias, heavily relies on the accurate interpretation of intracardiac signals \citep{huang2019catheter}. Additionally, cellular mechanisms reflected on these intracardiac recordings may reveal risk for cardiomyopathies \citep{rogers2021machine}. Nevertheless, the availability of databases containing intracardiac signals for these conditions remains exceedingly scarce. Additionally, the complexity of these signals and their sensitivity to noise present substantial obstacles, frequently hindering effective treatment \citep{starreveld2020impact}. 

Common traditional denoising methods such as template matching \citep{houben2006automatic}, beat averaging \citep{ng2006effect} and bandpass filtering \citep{venkatachalam2011signals}, fall short when dealing with the various noise sources \citep{de2022critical, stevenson2005recording}. In particular, electrophysiological (EP) noise, which arises from various factors like patient movement, electronic interference from medical devices, and physiological variations, becomes especially challenging due to its non-linear and non-stationary nature, rendering conventional denoising methods ineffective and demanding the use of an alternative approach \citep{starreveld2020impact}. To address this challenge, we propose the use of Variational Autoencoders (VAEs) \citep{kingma2013auto}. We hypothesize that VAEs have the potential to learn and interpret physiological morphologies of signals in this context, providing a novel solution to eliminate EP noise outperforming existing filtering techniques.

We develop a model capable of identifying and mitigating EP noise effects in easily verifiable clinically, yet challenging-to-obtain signal types. Our model, along with a unique dataset, presents a versatile denoising method that can be applied to less clinically verifiable signals, potentially advancing cardiac care and improving the outcomes of ablation therapy.

\section{Methods}

\begin{figure}[t]
\centering
\fbox{\includegraphics[width=1.0 \textwidth]{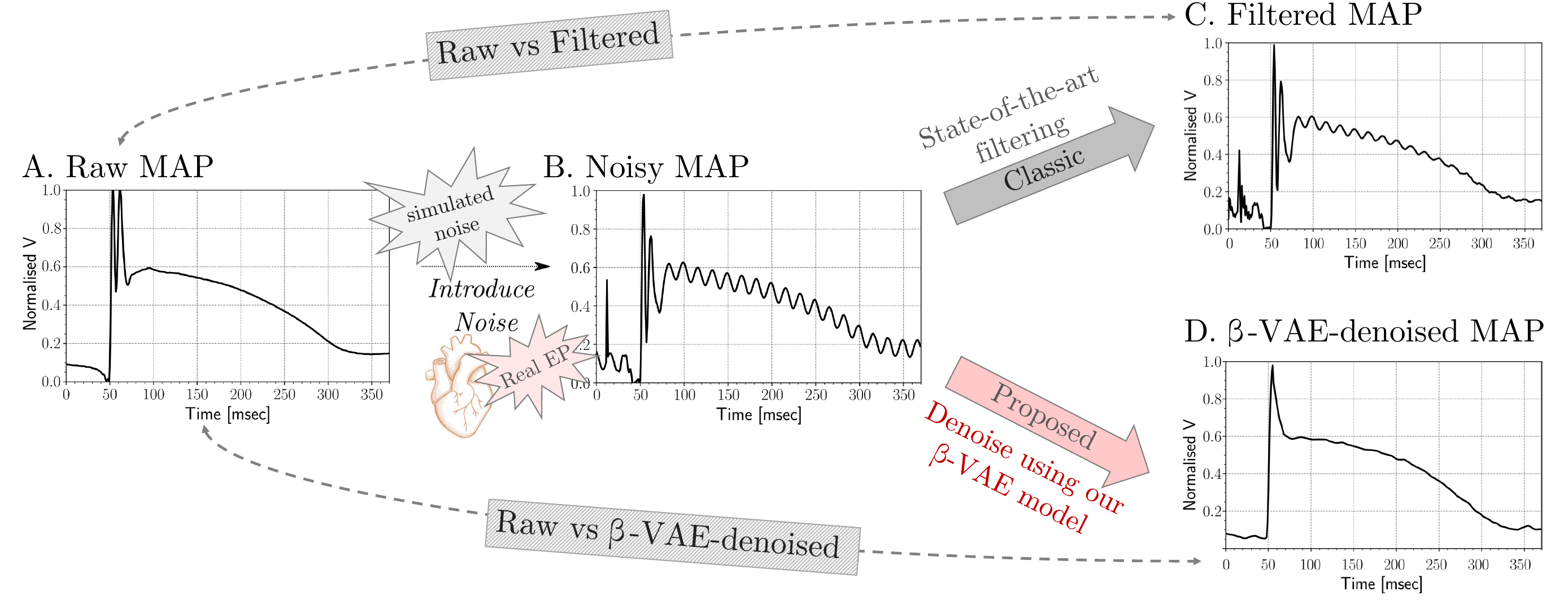}}
\caption[Abstract Figure]{{\footnotesize Denoising Pipeline. \textbf{A.} Raw intracardiac MAP recording from a patient with ischemic cardiomyopathy. \textbf{B.} Noisy counterpart of (A), following the introduction of various sources of simulated noise and EP noise. \textbf{C.} De-noised MAP time series obtained after applying the best-performing filter. \textbf{D.} Denoised MAP time series using our proposed $\beta$-VAE method.}}
\label{fig:denoising}
\end{figure}

\subsection{Denoising VAE Model}
Denoising Autoencoders (AEs) \citep{vincent2008extracting} are a type of encoder-decoder neural network that removes noise from corrupted or noisy data by learning to reconstruct the initial data $\mathbf{x}$ from its noisy counterpart $\tilde{\mathbf{x}}$. With \textit{g} as the encoding function and \textit{h} as the decoding counterpart, the primary objective becomes $\mathcal{L}_{D A E}=\left\|\mathbf{x}-h_{\varphi}\left(g_\theta(\tilde{\mathbf{x}})\right)\right\|^2$.

VAEs \citep{kingma2013auto} are a class of generative models designed to learn a probabilistic mapping between observed data and latent variables. Differing from traditional AEs, VAEs explicitly model the underlying data distribution using a generative model $p(\mathbf{x}, \mathbf{z})$, where $\mathbf{z}$ is the latent representation of the input. The ultimate goal of VAEs is to maximize the marginal likelihood $p_\theta(\mathbf{x})$,  which quantifies the model's effectiveness in explaining the observed data. To implement the VAE denoising model, we aim to maximize the tractable evidence lower bound (ELBO), decomposed in the Kullback-Leibler ($\mathbb{KL}$) divergence and the expected log-likelihood. The $\mathbb{KL}$ divergence quantifies the loss of information when approximating a posterior distribution to a simpler prior, typically a standard normal distribution. Additionally, it acts as a regularization term by encouraging similarity between these distributions (\cite{kingma2013auto}). On the other hand, when assuming a Gaussian likelihood with mean $\boldsymbol{\mu}$ and variance $\boldsymbol{\sigma}^2$
, the expected log-likelihood reduces to the expectation of the mean squared error (MSE) between $\mathbf{x}$ and $\mu(\mathbf{z})$ scaled up by a constant. We use Monte Carlo sampling to approximate such expectation. Lastly, following \cite{higgins2016beta}, we augment the original VAE framework with a $\beta$-weighting component to the $\mathbb{KL}$ term that modulates the learning constraints applied to the model. The ELBO becomes:
\begin{equation}
    \mathbf{E L B O}=\mathbb{E}_{q_{\varphi}(\mathbf{z} \mid \boldsymbol{x})}\left[\log p_\theta(\mathbf{x} \mid \mathbf{z})\right]-\beta \cdot \mathbb{KL} \left(q_{\varphi}(\mathbf{z} \mid \boldsymbol{x}) \| p_\theta(\mathbf{z})\right). 
\end{equation} 

For further details on the architecture of the VAE and its formulation, we refer to \Cref{sec:app_architecture,sec:app_vaemethod,sec:app_kl_twogaussians}.

\subsection{Noise Library}
Given the lack of ground truth clean MAP signals for noisy recordings, we conducted simulations to replicate prevalent forms of interference present in intracardiac signals. These include white noise, baseline wander, powerline interference, spike and truncation artifacts, as well as \textit{semi-synthetic }EP noise extracted from real physiological recordings. These interference patterns were later introduced into the recognizable MAPs -- see \cref{fig:denoising}.A.

We modeled Gaussian noise as a stationary and ergodic random process with zero mean and an autocorrelation function solely dependent on time lag. For baseline wander we combined a set of sinusoidal waves spanning a randomized range of low frequencies, from 0.01 to 0.3 Hz. To simulate powerline interference, we employed a single sinusoidal time series at 50Hz, adjusting the amplitude to mimic interference generated by various electrical devices and inadequate shielding. We introduced spike artifacts as finite discrete Dirac $\delta$-functions of varying amplitude, occurring within the initial segment of the MAP morphology -- before the upstroke. Additionally, we approximated truncation artifacts as multiplicative noise, using a square function with values set to one over the portion of the signal intended for retention while zeroing out the remainder. 

The computation of EP noise involved extracting noise from clinical recordings, operating on the premise that there is minimal variability in the true MAP morphology within a single patient. This is based on the understanding that MAP signals deviating significantly from the average recorded signal are likely to encompass noise stemming from diverse sources. As a result, we constructed a library of physiological noise time series extracted from patients' signals. These were subsequently merged to generate distinctive additive EP noise time series -- see an illustration in \cref{fig:denoising}.B.

\section{Experimental Settings}

\paragraph{Subject Recruitment and Study Protocol}
The study involved 53 individuals with low ventricular ejection fraction ($\leq40\%$) and coronary artery complications, undergoing ventricular stimulation. After exclusions, 42 patients remained. Conscious sedation was administered, and MAP intracardiac signals were recorded. Signals were pre-processed, resulting in 5706 individual MAP time series. The analysis focused on voltage-time series of cardiac beats within 370 ms windows, following post-alignment and artifact removal \cite{alhusseini2020machine}. Further details, including summarized baseline characteristics for the patient population, are described in \cref{sec:app_materials}.

\paragraph{Baselines}

There is a notable lack of established benchmarks for denoising intracardiac signals, particularly MAPs, in both laboratory and clinical settings.  We have explored various classic signal processing filtering techniques commonly used in clinical practice \citep{starreveld2020impact, de2022critical}. Among these techniques, the 5th order Butterworth filter \citep{yusuf2020analysis} is often considered the standard choice \citep{de2022critical, venkatachalam2011signals}, providing a solid baseline with potential efficacy in our specific dataset. The N-th order continuous-time Butterworth filter is characterized by its transfer function $
    H(s)= \prod_{k=1}^N\left(1-\frac{s}{p_k}\right)^{-1}
$ with poles $p_k$ located on the negative half-plane on a circle of radius $\omega_c$, as described by $
    p_k \triangleq \omega_c e^{j \frac{\pi(2 k+N-1)}{2 N}}
$.

\paragraph{Evaluation Metrics}
We assessed the model's effectiveness from diverse angles. This included using Pearson’s Correlation Coefficient (PCC) to measure the linear relationship between the original and denoised signals (Eq. \ref{eq:pcc}); calculating the Root Mean Squared Error (RMSE) to assess time domain alignment (Eq. \ref{eq:rmse}); and measuring the Power Signal to Noise Ratio (PSNR) (Eq. \ref{eq:psnr}) to contrast signal power with noise power, as quantified by the Mean Squared Error (MSE) -- see \cref{sec:app_evmetrics}.

\section{Results}
\begin{wraptable}[9]{r}{0.6\textwidth}
\vspace{-1.3em}
\caption{De-noising Results to All Noise Types}
\label{tab:denoising_all}
\resizebox{0.6\textwidth}{!}{%
\begin{tabular}{lllll}
\multicolumn{1}{c}{\bf Set} & \multicolumn{1}{c}{\bf Labels} & \multicolumn{1}{c}{\bf RMSE (x10-3) ↓} & \multicolumn{1}{c}{\bf PCC ↑} & \multicolumn{1}{c}{\bf PSNR ↑} \\
\hline \\
Train   & Noisy            & 14.98 {\small ± 0.40} & 0.867 {\small ± 0.007} & 20.52 {\small ± 0.13} \\ 
        & Filtered         & 13.67 {\small ± 0.31} & 0.883 {\small ± 0.007} & 20.40 {\small ± 0.08} \\
        & \textbf{Ours}     & \textbf{\hspace{0.17cm}2.68 {\small ± 0.43}}  & \textbf{0.990 {\small ± 0.001}} & \textbf{27.79 {\small ± 0.72}} \\
Test    & Noisy            & 15.41 {\small ± 0.60} & 0.864 {\small ± 0.024} & 20.33 {\small ± 0.24} \\
        & Filtered         & 14.45 {\small ± 0.55} & 0.879 {\small ± 0.022} & 20.21 {\small ± 0.19} \\
        & \textbf{Ours}      & \textbf{\hspace{0.17cm}7.05 {\small ± 0.89}}  & \textbf{0.967 {\small ± 0.009}} & \textbf{22.91 {\small ± 0.51}} \\
\end{tabular}%
}
\end{wraptable}
Noise poses a challenge in understanding physiological time series and is not effectively removed by current techniques in the clinic \cite{de2022critical, starreveld2020impact}, commonly negatively impacting treatment outcomes in cardiomyopathies or arrhythmias \citep{de2019electrogram}.

Our study leverages a VAE model to mitigate diverse sources of noise in physiological signals of the highly prevalent ischemic cardiomyopathies. 

\begin{wrapfigure}[15]{l}{0.6\textwidth}
\vspace{-1.2em}
\centering
\fbox{\includegraphics[width=0.6\textwidth]{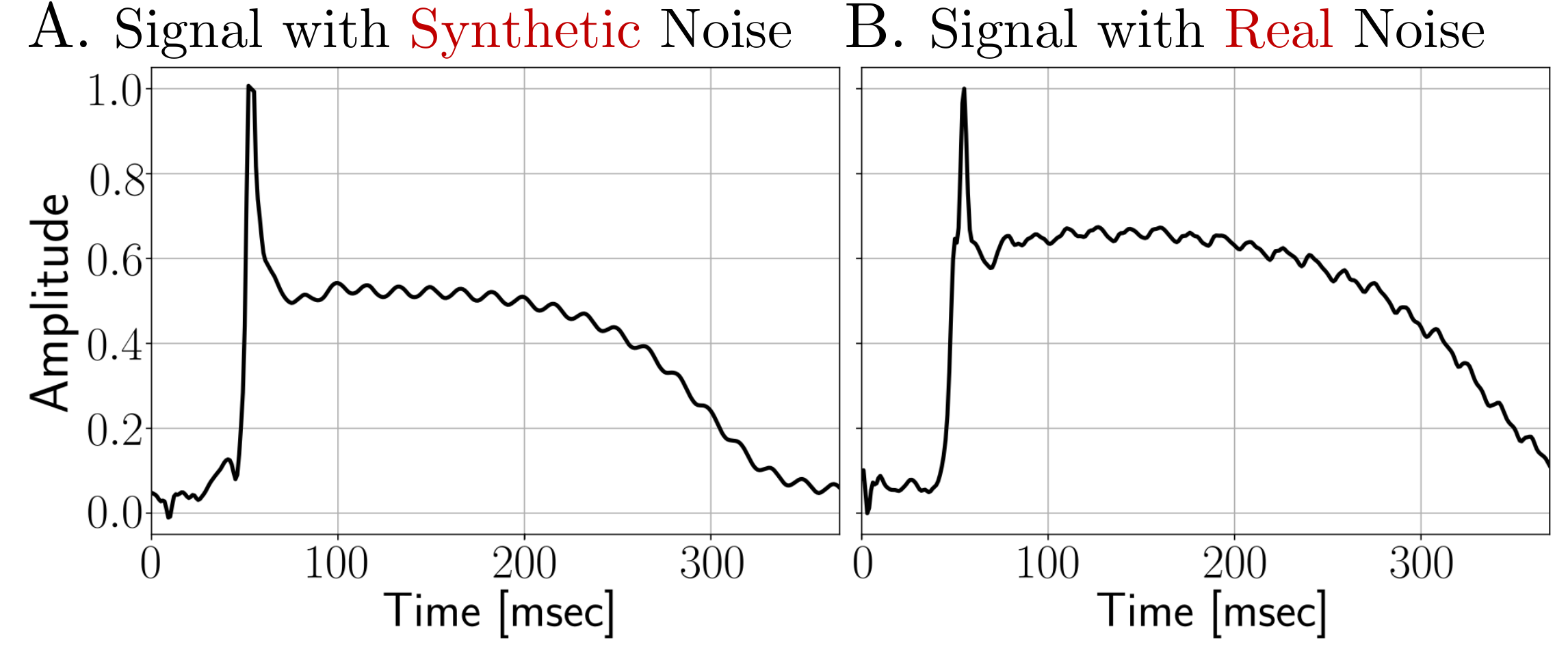}}
\caption[Synthetic Noise Reliability.]{{\footnotesize Synthetic Noise Reliability. \textbf{A.} Shows a raw MAP recording from an AF patient with added synthetic powerline interference noise. \textbf{B.} Presents a noisy recording of an MAP from a different patient, containing real powerline interference noise.}}
\label{fig:noise_reliability}
\end{wrapfigure}
\Cref{tab:denoising_noep} (in \Cref{sec:app_supplementary_results}) provides an overview of the evaluation metrics for denoising all noise types except EP noise, and \cref{tab:denoising_all} includes EP noise among the other types. These metrics compare the denoised MAP signals using our proposed $\beta$-VAE model, the benchmark (filtered) signals, and the noisy signals against the original clean ones. We find that our model successfully reduces different types of noise in intracardiac signals. Notably, it is effective in reducing clinical EP noise, which is usually difficult to filter out using traditional methods \citep{de2019electrogram, starreveld2020impact}. The model's ability to lessen the impact of EP noise is confirmed by comparing the denoising performance to all noise types but EP noise (see supplementary \cref{tab:denoising_noep}) against the denoising performance to all noise types, including EP noise (see \cref{tab:denoising_all}), observing only a minor decrease in performance.

One innovative aspect of our $\beta$-VAE model is how it denoises medical time series by learning representations of clean signals (and cellular processes) and uses this knowledge to identify and remove variations caused by noise. In our study, we focus on using MAPs, which are easier to visually verify by clinicians compared to other intracardiac signals \citep{franz1991method}. Compared to traditional filtering methods, the $\beta$-VAE model exhibits better denoising performance, especially in handling EP noise, which is challenging for standard filters given its non-linear and non-stationary nature. \Cref{fig:denoising} illustrates an example of a clinically interpretable MAP signal, as recognized by experts, alongside its noisy counterpart affected by various synthetic noise types. \Cref{fig:denoising}.C-D illustrates the effectiveness of the proposed model against the best-performing clinical filter, showing that the VAE-reconstructed MAP signal not only removes synthetic noise but also eliminates non-physiological features from the original recording, such as fractionated activation upstrokes. Additionally, our study employed both synthetic and real noise in testing the model's capabilities. The differentiation between these two types of noise is crucial, as they present varying levels of complexity and realism in the context of EP signal interference. 

Looking ahead, our research highlights several areas for further exploration. Bridging the domain gap between synthetic and real noise is important to explore further -- yet, \cref{fig:noise_reliability} indicates a strong visual resemblance between a synthetic noisy MAP and actual recordings with real noise. On the ethical aspect, while our denoising method holds promise for improving diagnostic accuracy and treatment outcomes, we must pay careful attention to potential risks and biases introduced. Developing clinical metrics to assess denoising effectiveness would be key in this regard. Additionally, validating the effectiveness of such models across different conditions and patient groups may be key for a broader applicability, especially in underserved communities. Moreover, exploring the latent space of these models will offer deeper insights into their functioning and potential improvements. Ultimately, our study sets the stage for crucial clinical validation in real-time settings, assessing the impact on patient outcomes and transitioning from theoretical models to practical medical solutions centered on patients. However, the responsible integration of AI technologies into healthcare workflows necessitates ongoing evaluation and validation to ensure patient safety and equitable access to care, something that will need to be taken into consideration moving forward.

\section{Conclusions}
Our study presents a VAE model designed to denoise intracardiac MAP signals, tackling the drawbacks of usual filtering methods used in the EP laboratory and clinical settings. This model, trained on a complex and elusive dataset mimicking clinical diagnostic scenarios, displays superior performance and resilience against different types of noise. Its skill in recreating important relevant features suggests it could be a big step forward for integration in real-time heart care settings, marking a significant advancement in cardiac care and improving the outcomes of EP therapies.

\bibliography{iclr2024_conference}
\bibliographystyle{iclr2024_conference}

\clearpage

\appendix
\section{Supplementary Results}
\label{sec:app_supplementary_results}
Additionally, Table \ref{tab:denoising_noep} offers an overview of the evaluation metrics for denoising all noise types except EP noise, complementing the information provided in Table \ref{tab:denoising_all}. 

\begin{table}[ht]
\caption{De-noising results to all noise sources but EP}
\label{tab:denoising_noep}
\begin{center}
\begin{tabular}{lllll}
\multicolumn{1}{c}{\bf Set} & \multicolumn{1}{c}{\bf Labels} & \multicolumn{1}{c}{\bf RMSE (x10-3) ↓} & \multicolumn{1}{c}{\bf PCC ↑} & \multicolumn{1}{c}{\bf PSNR ↑} \\
\hline \\
Train   & $\beta$-VAE      & \hspace{0.17cm}\textbf{2.18 {\small ± 0.44}}  &\textbf{0.992 {\small ± 0.001}} & \textbf{28.61 {\small ± 0.86}} \\
        & Filtered         & 11.76 {\small ± 0.27} & 0.899 {\small ± 0.005} & 21.09 {\small ± 0.13} \\
        & Noisy            & 12.88 {\small ± 0.20} & 0.884 {\small ± 0.006} & 21.34 {\small ± 0.13} \\
Test    & $\beta$-VAE      & \hspace{0.17cm}\textbf{6.25 {\small ± 0.86}}  &\textbf{0.970 {\small ± 0.009}} & \textbf{23.39 {\small ± 0.59}} \\
        & Filtered         & 12.09 {\small ± 0.63} & 0.899 {\small ± 0.017} & 21.04 {\small ± 0.21} \\
        & Noisy            & 12.91 {\small ± 0.80} & 0.885 {\small ± 0.018} & 21.30 {\small ± 0.29} \\
\end{tabular}
\end{center}
\end{table}

This comparison allows for an assessment of the model's generalizability in denoising EP noise.

In addition, \Cref{fig:noise_reliability} illustrates the strong visual resemblance between a synthetically constructed MAP signal with added synthetic noise, and a noisy signal as recorded directly from the device, featuring real noise.

\section{Supplementary Methods}

\subsection{Expanded Subject Recruitment and Study Protocol}
\label{sec:app_materials}
The research study protocol was approved by the Ethics Committee of Stanford University. The research protocol encompassed a cohort of 53 individuals who exhibited low ventricular ejection fraction ($\leq$40\%) along with coronary artery complications, scheduled for ventricular stimulation. Those individuals with a history of prolonged ventricular arrhythmias, sudden cardiac arrest, or fewer than 50 reliable monophasic action potential (MAP) recordings due to technical inconveniences were excluded from the study, reducing the population to 42 patients. The study was conducted under conscious sedation, utilizing midazolam and fentanyl. Specialized intracardiac signals were recorded using a dedicated 7F MAP catheter from Boston Scientific, MA, in conjunction with standard catheters. The high-fidelity MAP catheter was introduced either transvenously or through the aorta into the ventricles, with signals recorded using a physiological recorder from Bard-Boston Scientific in Marlborough, MA. These signals were filtered at various frequencies and digitized at a rate of 1 kHz. In brief, the study employed a total of 5706 individual MAP EGM time series exported at a 16-bit digital resolution. The analysis centered on voltage-time series of cardiac beats, which were partitioned within 370 ms windows obtained following alignment and the removal of artifactual deviations. Table \ref{tab:baselinechar} outlines the baseline characteristics of the entire patient population.

\begin{table}[ht]
\caption{Summary Baselines Characteristics of the Patient Population}
\label{tab:baselinechar}
\begin{center}
\begin{tabular}{ll}
\multicolumn{1}{c}{\bf Variable} & \multicolumn{1}{c}{\bf All Subjects (n=42)} \\ 
\hline \\
Age, y                            & 64.7 ± 13.0 \\
LVEF a, \%                        & 27.0 ± 7.6 \\
QRS Duration, ms                  & 126 ± 33 \\
Hypertension, \% (n)              & 19.0 (8) \\
Diabetes Mellitus, \% (n)         & 14.3 (6) \\
BNP b, pg/ml (median, IQR)        & 341 (157-999) \\
\end{tabular}
\end{center}
\end{table}

\subsection{Network Architecture}
\label{sec:app_architecture}
The encoder module processes input data (batch size = 32, input dim = 370) through six 1D convolutional layers, followed by flattening and two linear layers to produce mean and log variance of the latent distribution, with Leaky-ReLU activations after each layer. The decoder module takes a latent representation (batch size = 32, latent size = 32) and reconstructs the original input using an initial linear layer and a series of 1D transposed convolutional layers, mirroring the encoder's architecture, with similar post-layer activations.

\subsection{Denoising VAE Method}
\label{sec:app_vaemethod}
VAEs (\cite{kingma2013auto}) are a class of generative models designed to learn a probabilistic mapping between observed data and latent variables. Differing from traditional AEs, VAEs explicitly model the underlying data distribution using a generative model $p(\mathbf{x}, \mathbf{z})$. The ultimate goal of VAEs is to maximize the marginal likelihood $p_\theta(\mathbf{x})$,  which quantifies the model's effectiveness in explaining the observed data. To implement the VAE denoising model, we aim to maximize the tractable evidence lower bound (ELBO), decomposed in the Kullback-Leibler ($\mathbb{KL}$) divergence and the expected log-likelihood. The $\mathbb{KL}$ divergence quantifies the loss of information when approximating a posterior distribution to a simpler prior, typically a standard normal distribution. Additionally, it acts as a regularization term by encouraging similarity between these distributions (\cite{kingma2013auto}). Assuming $ q_{\varphi}\left(\mathbf{z} \mid \mathbf{x}^{(i)}\right)=\mathcal{N}\left(\mu, \operatorname{diag}\left\{\sigma_1^2, \ldots, \sigma_d^2\right\}\right) $ and a standard normal distribution for the prior, the $\mathbb{KL}$ divergence is derived as: 

\begin{equation}
    \mathbb{KL}\left(q_{\varphi}(\boldsymbol{z} \mid \boldsymbol{x}) \| p_\theta(\boldsymbol{z})\right)=\frac{\sum_{d=1}^D\left(1-\mu_d^2+\log \left({\sigma_d}^2\right)-\sigma_d^2\right)}{2}, 
\end{equation} 

see \Cref{sec:app_kl_twogaussians} for further detail.

Assuming the likelihood is Gaussian with mean $\mu$ and variance $\sigma^2$, the expected log-likelihood is expressed as:
\begin{equation}
    \log p_\theta(\boldsymbol{x} \mid \boldsymbol{z})=-\frac{\|\mathbf{x}-\mu(\mathbf{z})\|^2+\sigma^2 D \log \left(2 \pi \sigma^2\right)}{2 \sigma^2},
\end{equation}

which reduces to the expectation of the mean squared error (MSE) between $\mathbf{x}$ and $\mu(\mathbf{x})$ scaled up by a constant. 

Using Monte Carlo sampling to approximate such expectation, sampling $\Lambda$ latent variables $\mathbf{z}(\lambda)$ from $q_{\varphi}(\boldsymbol{z} \mid \boldsymbol{x})$, we find that:

\begin{equation}
 \mathbb{E}_{q_{\varphi}(\mathbf{z} \mid \mathbf{x})}\left[\log p_\theta(\mathbf{x} \mid \mathbf{z})\right] \approx \frac{\sum_{\lambda=1}^{\Lambda} \log p_\theta\left(\mathbf{x} \mid \mathbf{z}^\lambda\right)}{\Lambda}   
\end{equation}
    
Lastly, following \cite{higgins2016beta}, we augment the original VAE framework with a $\beta$-weighting component to the $\mathbb{KL}$ term that modulates the learning constraints applied to the model. Hence, we define the objective as:

\begin{equation}
    \mathbf{E L B O}=\mathbb{E}_{q_{\varphi}(\mathbf{z} \mid \boldsymbol{x})}\left[\log p_\theta(\mathbf{x} \mid \mathbf{z})\right]-\beta \cdot \mathbb{KL} \left(q_{\varphi}(\mathbf{z} \mid \boldsymbol{x}) \| p_\theta(\mathbf{z})\right). 
\end{equation}

\subsection{KL Divergence between two Gaussians and Specifics for our Prior}

\label{sec:app_kl_twogaussians}

Given two Gaussian distributions 
\[
q_\phi(\mathbf{z} | \mathbf{x}^{(i)}) = \mathcal{N}\left(\mu, \operatorname{diag}\left\{\sigma_1^2, \ldots, \sigma_D^2\right\}\right)
\]
and 
\[
p_\theta(\mathbf{z}) = \mathcal{N}(0, \textbf{I}),
\]
we want to compute the $\mathbb{KL}$ divergence between them.

The general formula for the $\mathbb{KL}$ divergence between two Gaussian distributions \( \mathcal{N}(\mu_1, \Sigma_1) \) and \( \mathcal{N}(\mu_2, \Sigma_2) \) is:
\begin{align}
\mathbb{KL}\left( \mathcal{N}(\mu_1, \Sigma_1) \,||\, \mathcal{N}(\mu_2, \Sigma_2) \right) &= \frac{1}{2} \Bigg( \text{Tr}(\Sigma_2^{-1} \Sigma_1) \nonumber + (\mu_2 - \mu_1)^T \Sigma_2^{-1} (\mu_2 - \mu_1) \nonumber \\
&- D - \log \frac{\det \Sigma_1}{\det \Sigma_2} \Bigg)
\end{align}

For our specific case, we have \( \mu_1 = \mu \), \( \Sigma_1 = \operatorname{diag}\{\sigma_1^2, \ldots, \sigma_D^2\} \), \( \mu_2 = 0 \), and \( \Sigma_2 = \textbf{I} \).

Substituting these into the formula, we get:
\begin{align}
\mathbb{KL}\left( q_\phi(\mathbf{z} | \mathbf{x}^{(i)}) \,||\, p_\theta(\mathbf{z}) \right) &= \frac{1}{2} \Bigg( \text{Tr}(\textbf{I}^{-1} \operatorname{diag}\{\sigma_1^2, \ldots, \sigma_D^2\}) \nonumber \\
&+ (0 - \mu)^T \textbf{I}^{-1} (0 - \mu) \nonumber \\
&- D - \log \frac{\det \operatorname{diag}\{\sigma_1^2, \ldots, \sigma_D^2\}}{\det \textbf{I}} \Bigg) \\
&= \frac{1}{2} \sum_{d=1}^D\left(1+\log \left(\left(\sigma_d\right)^2\right)-\left(\mu_d\right)^2-\left(\sigma_d\right)^2\right)
\end{align}

\subsection{Evaluation Metrics}
\label{sec:app_evmetrics}
The implementation of the evaluation metrics was performed as described by the equations below, including PCC to measure the linear relationship between the original and denoised signals:
\begin{equation}
\label{eq:pcc}
    \operatorname{PCC}_{s_1, s_2}=\frac{\sum_{i=1}^n\left(s_{1_i}-\overline{s_1}\right)\left(s_{2_i}-\overline{s_2}\right)}{\sqrt{\sum_{i=1}^n\left(s_{1_i}-\overline{s_1}\right)^2} \sqrt{\sum_{i=1}^n\left(s_{2_i}-\overline{s_2}\right)^2}}
\end{equation}
where \( s_1 \) and \( s_2 \) represent the raw MAP and its denoised counterpart, \( n \) denotes the number of data points in the signals, $i$ refers to the sample number within the respective signals, and \( \overline{s_1} \) and \( \overline{s_2} \) represent the mean values of \( s_1 \) and \( s_2 \) respectively.

The RMSE allowed us to assess the time domain alignment and sample-wise similarity of both the raw signal and denoised counterpart, implemented as:
\begin{equation}
\label{eq:rmse}
    \operatorname{RMSE}_{s_1, s_2}=\sqrt{\frac{\sum_{i=1}^n\left(s_{1_i}-s_{2_i}\right)^2}{n}}
\end{equation}
where \( s_1 \) and \( s_2 \) represent the signals being compared, \( n \) denotes the number of data points in the signals, and \( s_{1_i} \) and \( s_{2_i} \) represent individual data points within the respective signals.

Lastly, the PSNR permitted contrasting signal power to noise power, the latter quantified by the MSE:
\begin{equation}
\label{eq:psnr}
    \text { PSNR }=20 \cdot \log _{10}\left(\mathrm{MAX}_s\right)-\log _{10}(\mathrm{MSE})
\end{equation}
where \( \mathrm{MAX}_s \) denotes the maximum possible pixel value of the signal (e.g., for images, this would typically be 255 for 8-bit images).

\end{document}

%% file: math_commands.tex
\usepackage{amsmath,amsfonts,bm}

\def\eqref#1{equation~\ref{#1}}

\def\1{\bm{1}}

\DeclareMathAlphabet{\mathsfit}{\encodingdefault}{\sfdefault}{m}{sl}
\SetMathAlphabet{\mathsfit}{bold}{\encodingdefault}{\sfdefault}{bx}{n}













%% file: iclr2024_conference.bbl
\begin{thebibliography}{17}
\providecommand{\natexlab}[1]{#1}
\providecommand{\url}[1]{\texttt{#1}}
\expandafter\ifx\csname urlstyle\endcsname\relax
  \providecommand{\doi}[1]{doi: #1}\else
  \providecommand{\doi}{doi: \begingroup \urlstyle{rm}\Url}\fi

\bibitem[Alhusseini et~al.(2020)Alhusseini, Abuzaid, Rogers, Zaman, Baykaner, Clopton, Bailis, Zaharia, Wang, Rappel, et~al.]{alhusseini2020machine}
Mahmood~I Alhusseini, Firas Abuzaid, Albert~J Rogers, Junaid~AB Zaman, Tina Baykaner, Paul Clopton, Peter Bailis, Matei Zaharia, Paul~J Wang, Wouter-Jan Rappel, et~al.
\newblock Machine learning to classify intracardiac electrical patterns during atrial fibrillation: machine learning of atrial fibrillation.
\newblock \emph{Circulation: Arrhythmia and Electrophysiology}, 13\penalty0 (8):\penalty0 e008160, 2020.

\bibitem[Benjamin et~al.(2018)Benjamin, Virani, Callaway, Chamberlain, Chang, Cheng, Chiuve, Cushman, Delling, Deo, et~al.]{benjamin2018heart}
Emelia~J Benjamin, Salim~S Virani, Clifton~W Callaway, Alanna~M Chamberlain, Alexander~R Chang, Susan Cheng, Stephanie~E Chiuve, Mary Cushman, Francesca~N Delling, Rajat Deo, et~al.
\newblock Heart disease and stroke statistics—2018 update: a report from the american heart association.
\newblock \emph{Circulation}, 137\penalty0 (12):\penalty0 e67--e492, 2018.

\bibitem[de~Bakker(2019)]{de2019electrogram}
Jacques~MT de~Bakker.
\newblock Electrogram recording and analyzing techniques to optimize selection of target sites for ablation of cardiac arrhythmias.
\newblock \emph{Pacing and Clinical Electrophysiology}, 42\penalty0 (12):\penalty0 1503--1516, 2019.

\bibitem[De~Groot et~al.(2022)De~Groot, Shah, Boyle, Anter, Clifford, Deisenhofer, Deneke, Van~Dessel, Doessel, Dilaveris, et~al.]{de2022critical}
Natasja~MS De~Groot, Dipen Shah, Patrick~M Boyle, Elad Anter, Gari~D Clifford, Isabel Deisenhofer, Thomas Deneke, Pascal Van~Dessel, Olaf Doessel, Polychronis Dilaveris, et~al.
\newblock Critical appraisal of technologies to assess electrical activity during atrial fibrillation: a position paper from the european heart rhythm association and european society of cardiology working group on ecardiology in collaboration with the heart rhythm society, asia pacific heart rhythm society, latin american heart rhythm society and computing in cardiology.
\newblock \emph{EP Europace}, 24\penalty0 (2):\penalty0 313--330, 2022.

\bibitem[Franz(1991)]{franz1991method}
Michael~R Franz.
\newblock Method and theory of monophasic action potential recording.
\newblock \emph{Progress in cardiovascular diseases}, 33\penalty0 (6):\penalty0 347--368, 1991.

\bibitem[Gopinathannair et~al.(2015)Gopinathannair, Etheridge, Marchlinski, Spinale, Lakkireddy, and Olshansky]{gopinathannair2015arrhythmia}
Rakesh Gopinathannair, Susan~P Etheridge, Francis~E Marchlinski, Francis~G Spinale, Dhanunjaya Lakkireddy, and Brian Olshansky.
\newblock Arrhythmia-induced cardiomyopathies: mechanisms, recognition, and management.
\newblock \emph{Journal of the American College of Cardiology}, 66\penalty0 (15):\penalty0 1714--1728, 2015.

\bibitem[Higgins et~al.(2016)Higgins, Matthey, Pal, Burgess, Glorot, Botvinick, Mohamed, and Lerchner]{higgins2016beta}
Irina Higgins, Loic Matthey, Arka Pal, Christopher Burgess, Xavier Glorot, Matthew Botvinick, Shakir Mohamed, and Alexander Lerchner.
\newblock beta-vae: Learning basic visual concepts with a constrained variational framework.
\newblock In \emph{International conference on learning representations}, 2016.

\bibitem[Houben et~al.(2006)Houben, de~Groot, Lindemans, and Allessie]{houben2006automatic}
Richard~PM Houben, Natasja~MS de~Groot, Fred~W Lindemans, and Maurits~A Allessie.
\newblock Automatic mapping of human atrial fibrillation by template matching.
\newblock \emph{Heart Rhythm}, 3\penalty0 (10):\penalty0 1221--1228, 2006.

\bibitem[Huang \& Miller(2019)Huang and Miller]{huang2019catheter}
Shoei K~Stephen Huang and John~M Miller.
\newblock \emph{Catheter Ablation of Cardiac Arrhythmias E-Book: Catheter Ablation of Cardiac Arrhythmias E-Book}.
\newblock Elsevier Health Sciences, 2019.

\bibitem[Kingma \& Welling(2013)Kingma and Welling]{kingma2013auto}
Diederik~P Kingma and Max Welling.
\newblock Auto-encoding variational bayes.
\newblock \emph{arXiv preprint arXiv:1312.6114}, 2013.

\bibitem[Ng et~al.(2006)Ng, Kadish, and Goldberger]{ng2006effect}
Jason Ng, Alan~H Kadish, and Jeffrey~J Goldberger.
\newblock Effect of electrogram characteristics on the relationship of dominant frequency to atrial activation rate in atrial fibrillation.
\newblock \emph{Heart Rhythm}, 3\penalty0 (11):\penalty0 1295--1305, 2006.

\bibitem[Rogers et~al.(2021)Rogers, Selvalingam, Alhusseini, Krummen, Corrado, Abuzaid, Baykaner, Meyer, Clopton, Giles, et~al.]{rogers2021machine}
Albert~J Rogers, Anojan Selvalingam, Mahmood~I Alhusseini, David~E Krummen, Cesare Corrado, Firas Abuzaid, Tina Baykaner, Christian Meyer, Paul Clopton, Wayne Giles, et~al.
\newblock Machine learned cellular phenotypes in cardiomyopathy predict sudden death.
\newblock \emph{Circulation Research}, 128\penalty0 (2):\penalty0 172--184, 2021.

\bibitem[Starreveld et~al.(2020)Starreveld, Knops, Roos-Serote, Kik, Bogers, Brundel, and de~Groot]{starreveld2020impact}
Roeliene Starreveld, Paul Knops, Maarten Roos-Serote, Charles Kik, Ad~JJC Bogers, Bianca~JJM Brundel, and Natasja~MS de~Groot.
\newblock The impact of filter settings on morphology of unipolar fibrillation potentials.
\newblock \emph{Journal of Cardiovascular Translational Research}, 13:\penalty0 953--964, 2020.

\bibitem[Stevenson \& Soejima(2005)Stevenson and Soejima]{stevenson2005recording}
William~G Stevenson and Kyoko Soejima.
\newblock Recording techniques for clinical electrophysiology.
\newblock \emph{Journal of cardiovascular electrophysiology}, 16\penalty0 (9):\penalty0 1017--1022, 2005.

\bibitem[Venkatachalam et~al.(2011)Venkatachalam, Herbrandson, and Asirvatham]{venkatachalam2011signals}
KL~Venkatachalam, Joel~E Herbrandson, and Samuel~J Asirvatham.
\newblock Signals and signal processing for the electrophysiologist: part ii: signal processing and artifact.
\newblock \emph{Circulation: Arrhythmia and Electrophysiology}, 4\penalty0 (6):\penalty0 974--981, 2011.

\bibitem[Vincent et~al.(2008)Vincent, Larochelle, Bengio, and Manzagol]{vincent2008extracting}
Pascal Vincent, Hugo Larochelle, Yoshua Bengio, and Pierre-Antoine Manzagol.
\newblock Extracting and composing robust features with denoising autoencoders.
\newblock In \emph{Proceedings of the 25th international conference on Machine learning}, pp.\  1096--1103, 2008.

\bibitem[Yusuf et~al.(2020)Yusuf, Maduakolam, Umar, and Loko]{yusuf2020analysis}
Samson~D Yusuf, Francis~C Maduakolam, Ibrahim Umar, and Abdulmumini~Z Loko.
\newblock Analysis of butterworth filter for electrocardiogram de-noising using daubechies wavelets.
\newblock \emph{SSRG International Journal of Electronics and Communication Engineering}, 7\penalty0 (4):\penalty0 8--13, 2020.

\end{thebibliography}
